\documentstyle[12pt]{article}
\begin{document}
\newcommand{\beq}{\begin{equation}}
\newcommand{\eeq}{\end{equation}}
\newcommand{\beqn}{\begin{eqnarray}}
\newcommand{\eeqn}{\end{eqnarray}}
\newcommand{\slp}{\raise.15ex\hbox{$/$}\kern-.57em\hbox{$\partial$}}
\newcommand{\slA}{\raise.15ex\hbox{$/$}\kern-.57em\hbox{$A$}}
\newcommand{\slD}{\raise.15ex\hbox{$/$}\kern-.57em\hbox{$D$}}
\newcommand{\hf}{h_{k_{F}}}
\newcommand{\lnA}{\raise.15ex\hbox{$/$}\kern-.57em\hbox{$A$}}
\newcommand{\lnC}{\raise.15ex\hbox{$/$}\kern-.57em\hbox{$C$}}
\newcommand{\sls}{\raise.15ex\hbox{$/$}\kern-.57em\hbox{$s$}}
\newcommand{\bP}{\bar{\Psi}}
\newcommand{\bC}{\bar{\chi}}
\newcommand{\phiH}{\hat{\phi}}
\newcommand{\etaH}{\hat{\eta}}
\title{Friedel oscillations in a Luttinger liquid with long-range interactions}

\vspace{2cm}

\author{Victoria I. Fern\'andez$^{a, b}$ and Carlos M. Na\'on$^{a, b}$}

\date{Revised, February 2001}

\maketitle

\def\thepage{\protect\raisebox{0ex}{\ } La Plata-Th 00/12}
\thispagestyle{headings} \markright{\thepage}

\begin{abstract}
\indent We introduce a path-integral approach that allows to compute charge density
oscillations in a Luttinger liquid with impurities. We obtain an explicit expression
for the envelope of Friedel oscillations in the presence of arbitrary
electron-electron potentials. As examples, in order to illustrate the procedure, we
show how to use our formula for contact and Coulomb potentials.

\end{abstract}

\vspace{3cm}

\noindent Pacs: 05.30.Fk, 71.10.Pm\\


\noindent --------------------------------

\noindent $^a$ {\footnotesize Instituto de F\'{\i}sica La Plata, Departamento de
F\'{\i}sica, Facultad de Ciencias Exactas, Universidad Nacional de La Plata.  CC 67,
1900 La Plata, Argentina.}

\noindent $^b$ {\footnotesize CONICET.\\ e-mails: victoria@venus.fisica.unlp.edu.ar,~
naon@venus.fisica.unlp.edu.ar}

\newpage
\pagenumbering{arabic}

In the last years there has been much activity addressed to the study of condensed
matter and statistical mechanics problems through field-theoretical methods
\cite{Fradkin}. In particular the physics of one-dimensional (1d) systems of strongly
correlated particles has become a very interesting subject since one can take
advantage of the simplicity of the models at hand and, at the same time, expect to
make contact with experiments. For instance, the recently built quantum wire
\cite{quantum wires} is a good realization of a 1d electron gas. From the theoretical
side, the simplest formulation of a 1d electronic system is given by the
Tomonaga-Luttinger (TL) model \cite{TL} which has been successful in describing some
qualitative features of a Luttinger liquid such as spin-charge separation and
non-universal exponents in the decay law of correlation functions \cite{Voit}. There
are, however, two crucial issues that are not considered in the original versions of
this model: the presence of a non trivial interaction between electrons and impurities
\cite{KF} and the effect of long-range (LR) electron-electron interactions
\cite{Schulz} \cite{long range}. As it is well known, the former leads to the
occurrence of Friedel oscillations in the charge density profile, at least for Fermi
liquids \cite{Friedel}. On the other hand, as the dimensionality of a system
decreases, charge screening effects become less important and the LR interaction
between electrons is expected to play a central role in determining the properties of
the system. In fact, from a theoretical point of view, the effects of LR interactions
have been recently discussed in connection to several problems such as the Fermi-edge
singularity \cite{Fermi-edge}, the insulator-metal transition \cite{insulator-metal}
and the role of the lattice through umklapp scattering and size dependent effects
\cite{lattice}. Thus, it is quite interesting to study the interplay between
impurities and LR interactions by considering Friedel oscillations in a 1d system.
Some time ago, Egger and Grabert \cite{EG} analyzed this phenomenon. By combining the
techniques of standard bosonization \cite{conventional bosonization} with the
self-consistent harmonic approximation \cite{SCHA} and quantum Monte Carlo simulations
\cite{QMC}, they were able to get explicit results for both weak and strong impurity
scattering regimes. Later on, the authors of Ref. \cite{LLS} used bosonization and a
scattering description to get some exact results for the short-range case and for a
special value of the coupling constant, equivalent to the so called "Toulouse point"
in the anisotropic Kondo problem. More recently, the authors of Ref. \cite{YCZC} used
again standard bosonization to address the same problem emphasizing the equivalence
between the TL model in the presence of a single non-magnetic impurity and a boundary
Sine-Gordon model.\\
In this work we present an alternative approach to this problem based on a
path-integral bosonization technique previously developed in the context of non local
quantum field theories \cite{nonlocal bosonization}. This method seems to be specially
adequate to consider LR interactions. Indeed, it has recently provided a
straightforward derivation of the electronic Green's function in the presence of non
contact potentials \cite{Iucci}. Then, our main purpose here is to show how to extend
this formulation to the computation of Friedel oscillations.\\
\vspace{0.5cm}

We start from a modified non local Thirring model \cite{FLN} described by the
following (Euclidean) Lagrangian density:
\beq
{\cal L}=i\bar{\Psi}(\slp+\gamma_{0}k_F)\Psi + \int d^2 yJ_{\mu}(x)U_{(\mu )}(x,y)
J_{\mu}(y) +\bar{\Psi}\lnC\Psi - M(x)\bar{\Psi}\Psi
\label{1}
\eeq
where $x=(\tau_{x},{\bf x})=(x_0,x_1)$, and $J_{\mu}={\bP}\gamma_{\mu}{\Psi}$. The functions
$U_{(\mu)}(x,y)$ are forward-scattering potentials. Setting
$U_{(0)}=U_{(1)}=-\delta^2(x-y)$ one gets the covariant and local version of the
Thirring model usually studied in the context of (1+1) QFT's.\\
On the other hand, the
choice $U_{(0)}(x,y)=U(|{\bf x}-{\bf y}|) \,\ \delta(\tau_x-\tau_y)$ and $U_{(1)}(x,y)=0$,
yields the simplest version of the Tomonaga-Luttinger (TL) model with an instantaneous
distance dependent potential and
no current-current fluctuations. The last two terms in (\ref{1}) correspond to forward
and backward electron-impurity scattering, respectively.\\
The main purpose of the present paper is to evaluate the v.e.v. of the charge-density:
\beqn
\langle \rho (x) \rangle = \langle \Psi ^{\dagger} \Psi +
e^{-2i k_{F} {\bf x}} \Psi ^{\dag}_{R} \Psi_{L} + e^{2i k_{F}{\bf
x}}\Psi ^{\dag}_{L} \Psi_{R} \rangle,
\label{3}
\eeqn
for an arbitrary electron-electron potential $U_{(\mu)}(x,y)$. Also we will be specially
interested in case that the impurity terms are $C_0(x) = V \delta ({\bf x}-d)= M(x)$ and
$C_1(x)=0$, where V is a constant proportional to the impurity tunneling
barrier situated at ${\bf x}=d$. Using a suitable representation
of the functional delta and introducing an auxiliary vector field
$A_{\mu}$ (see Ref. \cite{nonlocal bosonization} for details), the
partition function of the model under consideration reads
\beq
Z = N \int {\cal D}A_{\mu}\,e^{-S[A]} det\left(i \slp + \sqrt{2}\, \lnA + \lnC + \sls +
i\gamma_{0}k_F-M(x)+s_{\mu} \epsilon_{\mu \nu} e^{2ik_{F}{\bf x}
\Gamma_{\nu}}\right)
\label{9}
\eeq
\noindent where  $S[A]$ is the free quadratic action for
$A_{\mu}$, $\Gamma_{0}=\mbox{$\bf{I}$}$ and $\Gamma_{1}=
\gamma_{5}$. Equation (\ref{3}) can be obtain by functional derivation of Eq.
(\ref{9}) with respect to the source $s_{0}$.\\
As it is known, the massive-like determinant in equation (\ref{9}) cannot be exactly solved,
even in the local case. However, we can take advantage of the fact that the vacuum to
vacuum functional can be written in such a way that non local terms are not present in
the determinant. Therefore the terms $ \sqrt{2}\, \lnA + \lnC + \sls + i\gamma_{0}k_F$
can be decoupled from fermions by performing chiral and gauge transformations in the
fermionic path-integral measure. Indeed, decomposing $A_{\mu}(x)$ in longitudinal and
transverse pieces
\beq
A_{\mu}(x)=\epsilon_{\mu\nu}\partial_{\nu}\left(\Phi (x)
-\frac{ik_{F}{\bf x}}{\sqrt{2}}\right) +
\partial_{\mu}\eta (x)- \frac{1}{\sqrt{2}}(C_{\mu}(x)+ s_{\mu}(x)),
\eeq
\noindent where $\Phi$ and $\eta$ are boson fields (to be associated to the
normal modes of the system) and applying, as anticipated, functional bosonization
techniques \cite{nonlocal bosonization} to express the fermionic
determinant in terms of $\Phi$ and $\eta$, one finally obtains
\begin{equation}
Z=N \int {\cal D}\bar\chi {\cal D}\chi {\cal{D}} \Phi {\cal{D}} \eta
\,\exp{(-S_{bos})}\,\exp{(-S_{fer})}\, \exp{(-S[M,s_{\mu}])},
\end{equation}
\noindent where $S_{fer}$ corresponds to free massless fermions ($\chi$ and
$\bar\chi$) and
\beq
S[M,s_{\mu}] =  \int d^2x\, \bar\chi\left(s_{\mu}(x)
\,\epsilon_{\mu \nu}\, e^{2ik_{F}{\bf x} \Gamma_{\nu}} - M(x)\right)
e^{-2g\gamma_5\Phi}\chi.
\eeq
Concerning $S_{bos}$, it can be more briefly described
in momentum space:
\beq
\begin{array}{cl}
 S_{bos}=&\int \frac{ d^2p }{(2\pi)^2}
           \pmatrix{\hat{\Phi}(p)&\hat{\eta}(p)&\hat{C'_{\mu}}(p)\cr}\pmatrix{A(p)&\frac{C(p)}{2}&\frac{E(p)}{2}\cr \frac{C(p)}{2}&B(p)&\frac{F(p)}{2}\cr \frac{E(p)}{2}&\frac{F(p)}{2}&D(p)\cr}\pmatrix{\hat{\Phi}(-p)\cr \hat{\eta}(-p)\cr \hat{C'_{\mu}}(-p)\cr}+\nonumber\\
&+\frac{ik_{F}}{2}\int \frac{d^2p}{(2\pi)^2}\hat{U}^{-1}_{(0)}(p)
\hat{C'_{(0)}}(p)\delta^2(p),
\end{array}
\eeq
\noindent where we have defined $C'_{\mu}=C_{\mu}+s_{\mu}$
and $A(p), B(p), C(p), D(p), E(p)$ and $F(p)$ are potential dependent functions (See
\cite{FLN} for more details).\\
At this point we see that the generating functional can be
formally expanded in powers of $\left(M- s_{\mu} \epsilon_{\mu
\nu} e^{2ik_{F}{\bf x} \Gamma_{\nu}}\right)$, in complete analogy
with the usual procedure employed in the path-integral
bosonization of $(1+1)$ massive QFT's \cite{massive bos}
\cite{LN}. In fact, the $x$ dependence of this perturbative
parameter, together with the appearance of $C'_{\mu}(x)$ in the
bosonic action are two of the new features of the present
computation. As far as these functions are well-behaved one can
assume the existence of every term in the corresponding series.\\
From now on we will specialize the computation to the case
$s_{1}=0$. This allows to define $M_{\pm}(x)= M(x)-s_{0}(x)e^{\mp
2ik_{F}{\bf x}}$ and one can then perform the above mentioned
expansion of $Z$ taking $M_{\pm}(x)$ as perturbative parameters.
As explained in Ref. \cite{FLN} one can show that the same
expansion can be obtained by starting from a purely bosonic non
local extension of the sine-Gordon model given by
\beqn
{\cal L'}&=& \frac{1}{2}(\partial_{\mu} \varphi)^2 + \frac{1}{2}\displaystyle\int d^2 y
\,\partial_{\mu}\varphi(x)\,d_{(\mu )}(x,y)\, \partial_{\mu}\varphi(y)+ \nonumber\\
&&+F_{\mu}\,\partial_{\mu}\varphi - \frac{1}{2 \beta^2}\left(\alpha_{+} (x)\,
e^{i \beta\varphi (x)} + \alpha_{-} (x)\,e^{-i \beta\varphi (x)}\right)
\label{26}
\eeqn
where $F_{\mu}(x)$ represents a couple of classical functions to
be related to the $C'_{\mu}$'s and $d_{(\mu )}(x,y)$ are two
bilocal functions that will be associated to the electron-electron
potentials (a similar non locality in the kinetic term was
considered in the study of the influence of LR correlations in the
metal-insulator transition \cite{Krive}). $\beta$ is a constant
and $\alpha_{\pm} (x)$ are functions that can be considered as
extensions of the parameter $\alpha_0$ used by Coleman
\cite{Coleman}. Indeed, for $d_{(\mu)}=0=F_{\mu}$ and $\alpha_{+}=
\alpha_{-}= \alpha_0= constant$, the model above coincides with
the usual sine-Gordon model. In the present approach the
quantities $\alpha_{\pm} (x)$ are related to $M_{\pm}(x)$, which
are in turn connected to the strength of the scatterer.
Let us stress that in our formulation it is
straightforward to consider a non point-like impurity
($\alpha_{\pm} (x) \neq V \delta ({\bf x}-d)$). However, in order
to illustrate our method, in this paper we will consider the usual
case of a completely localized impurity. For this particular case
Eq. (\ref{26}) contains the same terms that can be derived from
standard bosonization (See for instance \cite{EG}).

Now, going to momentum space and employing standard procedures to
evaluate each v.e.v., the partition function $Z'$ corresponding to
this generalized sine-Gordon model coincides with $Z$ provided
that the following three relations hold:

\beq
\frac{2\pi} {\frac{2}{\pi}[p_0^2 \hat{U}_{(1)}(p) +p_1^2\hat{U}_{(0)}(p)] + p^2}=
\frac{\beta^2} {2[p^2+ \hat{d}_{(0)}(p) p_0^2+ \hat{d}_{(1)}(p) p_1^2] },
\label{28}
\eeq

\beq
\frac{\alpha_{\pm} (x)}{\beta^2}= M_{\pm}(x)= M(x)-s_{0}(x)e^{\mp 2ik_{F}{\bf x}}
\label{29}
\eeq

\beq
\frac{2i \hat{C'}_{\mu}(-p)\epsilon_{\mu\nu} p_{\nu}}{\frac{2}{\pi}[p_0^2
\hat{U}_{(1)}(p) +p_1^2\hat{U}_{(0)}(p)] + p^2}=
\frac{\beta\hat{F_{\mu}}(-p)p_{\mu}}{p^2+ \hat{d}_{(0)} p_0^2+ \hat{d}_{(1)} p_1^2}.
\label{30}
\eeq
\\
\noindent Therefore, we have obtained an equivalence between the partition functions
$Z$ and $Z'$ corresponding to the  non local Thirring and sine-Gordon models with
extra interactions defined above. This means that we can use $Z'$ together with the
above conditions in order to compute the charge-density in the Luttinger liquid in the
presence of impurities. Indeed, as a result of this bosonization technique we can
evaluate $<\rho(x)>$ through functional derivation of $Z'$ instead of $Z$. In so doing we
obtain
\beqn
\langle \rho (x) \rangle = \langle \frac{i}{\sqrt{\pi}}
\partial_{{\bf x}} \varphi (x) + \cos \left(2\sqrt{\pi} \varphi (x) - 2 k_{F} {\bf x}\right)
\rangle
\label{31}
\eeqn
where the v.e.v. is taken with respect to the Lagrangian
density ${\cal L'}[s_{0}=0]$ obtained from (\ref{26}) after using equations
(\ref{28}), (\ref{29}) and (\ref{30}) and setting $s_{0}=0$. Note that we have also
set $\beta=2\sqrt{\pi}$.\\
Let us remark that there is an additional contribution to Eq.
 (\ref{31}), coming from the functional derivative of the normalization constant
$N'[C'_{\mu}]$ with respect to $s_{\mu}$. Since this quantity is a constant, its only
effect is to shift the background value of the charge density. For this reason we have
just disregarded it.\\
Now we return to our main goal, that is to use the path-integral
framework depicted above in order to obtain an explicit formula
for the charge-density in a Luttinger liquid with an arbitrary
electron-electron and electron-impurity interactions. When one
imposes these conditions in ${\cal L'}[s_{0}=0]$ one gets a
Lagrangian density which has an undefined parity as functional of
$\varphi$. However it is much simpler to work with an even
Lagrangian since in this case all v.e.v.'s of odd functions of
$\varphi$ will vanish. It is easy to see that the translation
$\varphi(x) \rightarrow \varphi(x) + f(x)$ yields an even
Lagrangian ${\cal L'}_{even}$ provided that the classical function
$f(x)$ is $\tau_{x}$-independent and its gradient satisfies:
\beq
\partial_{\bf x} f(x) +\frac{2}{\pi}\int d{\bf y}\,U({\bf x}-{\bf y})\,
\partial_{\bf y} f(\tau_{x},{\bf y}) + \frac{i}{\sqrt{\pi}}\,V\,\delta({\bf x}-d) = 0.
\label{35}
\eeq
\noindent We then get
\beq
\langle \rho (x) \rangle = \frac{i}{\sqrt{\pi}} \partial_{\bf x} f + \cos (\sqrt{
4 \pi} f({\bf x})-2 k_{F} {\bf x}) \langle \cos (\sqrt{4 \pi}\varphi(x) ) \rangle _
{{\cal L'}_{even}},
\label{36}
\eeq
\noindent where $\cos (\sqrt{ 4 \pi} f({\bf x})-2 k_{F} {\bf x})$ is called the
Friedel oscillation and $A(x)= \langle \cos (\sqrt{ 4 \pi}\varphi(x))
\rangle_{{\cal L'}_{even}}$ is the corresponding envelope.\\
Let us point out that Eq. (\ref{35}) has been previously found in Ref. \cite{YCZC}. As
shown by these authors, in the short-range case it has the solution $f=constant
\propto U$ whose only effect is to add a constant phase in the cosine term associated
to the Friedel oscillation. A non trivial phenomenon takes place for LR potentials,
since the cosine ceases to be a periodic function. This nonperiodicity effect,
although weak at large distances, could eventually be observed in carbon nanotubes
\cite{nanotubes}.\\
From now on we shall focus our attention on the computation of the envelope of the oscillation.
Since ${\cal L'}_{even}$ is not exactly solvable, we shall employ the well-known self-consistent
harmonic approximation \cite{SCHA}, which amounts to replacing ${\cal L'}_{even}$ by
\beq
{\cal L}_{SCHA} = \frac{1}{2}(\partial_{\mu}\varphi)^2+\frac{1}{\pi} \int d
{\bf y}\,
\partial_{\bf x} \varphi(\tau_{x},{\bf x})\, U({\bf x}-{\bf y})\,\partial_{\bf
y}\varphi(\tau_{x},{\bf y}) + \frac{m(V)\, \delta ({\bf x} - d)}{2}\, \varphi^2
\eeq
where $m(V)$ is a constant, related to the impurity strength, to be
variationally determined. The precise relationship between $m(V)$ and $V$ was obtained
in Ref. \cite{EG}. For instance, in the strong-scattering limit, when $V$ is much
larger than a certain bandwidth, one has simply $m=V$ (See also Ref. \cite{YCZC}).\\
Let us now consider the computation of $A(x)$ using this approximation. Performing a
translation in the field $\varphi(x) \rightarrow \varphi(x)+ a(x)$, with $a(x)$ a
classical function, we find $ A(x)= \exp{i \sqrt{\pi} a(x)}$.\\
Going to momentum space we see that the Fourier transform of $a(x)$ satisfies an integral
equation whose solution is
\beq
\hat{a}(p) = \frac{2i\sqrt{\pi}}{p^2+
p_{1}^{2}\frac{2U(p_{1})}{\pi}} e^{-i p_{\mu} x_{\mu}} \left( 1-
\frac{m e^{i p_{1} r}I ( p_{0}, r)}{\pi + m I ( p_{0}, 0)}\right)
\eeq
\noindent with
\beq
I ( p_{0}, r) = \int^{\infty}_{0} d q_{1} \frac{\cos( q_{1}
r)}{p_{0}^{2}+q_{1}^{2} (1 + \frac{2 U( q_{1})}{\pi})},
\label{43}
\eeq
where we defined $r=|{\bf x}-d|$.\\
The envelope of the Friedel oscillation then reads
\beq
A(r)= \exp{- \frac{1}{\pi} \int^{\infty}_{-\infty} d p_{0} \left( I ( p_{0}, 0)-
\frac{m I ^{2}( p_{0}, r)}{\pi + m I ( p_{0}, 0)}\right)},
\label{44}
\eeq
which is our main formal result. Indeed, formulae (\ref{43}) and (\ref{44}) give an analytical
expression (exact within the gaussian approximation) for $A(r)$ as functional of both
the electron-electron potential and the variational parameter $m(V)$. Since the
self-consistent harmonic approximation seems to fail for weak impurity strength, due
to the neglect of interwell tunneling \cite{EG}, we restrict our analysis to the
strong impurity regime. We also consider a large distance approximation of Eq.
(\ref{43}) which consists of inserting $1/r$ as infrared cutoff. Let us call
$I_{r}(p_0,r)$ the integral (\ref{43}) regulated in this way. We will examine, as
examples, two specific short range and Coulomb potentials. This, in turn will allow us
to illustrate how to use our general formula (\ref{44}) for other cases. Moreover,
since these problems were previously considered in Refs. \cite{EG} and \cite{YCZC} by
using standard (operational) bosonization, our computation will give an independent
confirmation by means of a different approach. First of all we note that it is
convenient to split out the two terms of the exponential factor on Eq. (\ref{44}), such that
$A(r)= \exp\left(T + W\right)(r)$.\\
For the simple contact potential $U(q_1)=U=constant$, we get
\beq
T(r)= \ln (\Lambda r)^{-g}
\eeq
and
\beq
W(r)= g\, \exp( mg^2 r)\,E_{i} (-mg^2 r)- g \,\exp( 2g)\, E_{i} ( -2g)
\eeq
\noindent where $\Lambda$ is an ultraviolet cutoff. We have also introduced the
interaction constant $g= (1 + \frac{2 U}{\pi})^{-\frac{1}{2}}$. Taking into account
the asymptotic behaviour of the exponential integral function $E_{i}$ for $mg^2 r
>> 1$, one obtains
\beq
A(r) = {\cal C}(g,\Lambda)\, (2g\,r)^{-g} \, \exp(- \frac{1}{m\, g\, r})
\eeq
\noindent which coincides with Refs. \cite{EG} and \cite{YCZC} under the same
regime.\\
In the Coulombian case, one has $U(|{\bf x}|)= U/\sqrt{|{\bf x}|^{2} + b^{2}}$, whose
Fourier transform is $U(q_{1})= 2\, U K_{0} (b\, q_{1})$, where $b$ plays the role of
a lattice spacing. Inserting this expression in $T$ and $W$, and considering the same
regime as before we find that $W$ vanishes and
\beq
T(r) = - \frac{\pi}{2 U}\left(\sqrt{1 - \frac{4 U}{\pi} \ln \frac{b}{2 r}}-
\sqrt{1 - \frac{4 U}{\pi} \ln{ \frac{b \Lambda}{2}}} \,\,\right)
\eeq
\noindent which yields
\beq
A(r) = {\cal C'}(g,b,\Lambda)  \exp \left(- \sqrt{ \frac{\pi}{U}\ln\,
\frac{r}{b}}\,\right).
\eeq
\noindent Again, this behaviour is equal to the one previously found in Refs.
\cite{EG} and \cite{YCZC}.\\
\vspace{0.75cm}

In summary, we have described an alternative bosonization approach to
the computation of charge-density fluctuations. This technique, previously originated
in the context of QFT's, parallels, in the path-integral framework, the operational
schemes usually employed in condensed matter applications. In particular, we have
computed the envelope of Friedel oscillations in a simple version of the TL model with
a non-magnetic impurity. By combining that bosonization procedure and the
self-consistent harmonic approximation we were able to express the envelope of the
oscillations as a functional of the electron-electron interaction (see Eqs. (\ref{43})
and (\ref{44})). Finally, as a consistency check of this formal result, and in order
to illustrate our method, we considered the long-distance regime for contact
interactions and Coulomb potentials. Our results are in agreement with Refs. \cite{EG}
and \cite{YCZC}.
\\
\\
\paragraph{Acknowledgement}
This work was partially supported by the Consejo Nacional de Investigaciones
Cient\'{\i}ficas y T\'ecnicas (CONICET), Argentina.\\ We are grateful to An\'{\i}bal
Iucci for useful discussions.

\end{document}